\documentstyle[pra,preprint,aps]{revtex}

\begin{document}
\bibliographystyle{prsty}
\title{Loading of a Rb magneto-optic trap from a getter source }
\author{Umakant D. Rapol, Ajay Wasan, and Vasant 
Natarajan\thanks{Electronic mail: vasant@physics.iisc.ernet.in}}
\address{Department of Physics, Indian Institute of Science, 
Bangalore 560 012, INDIA}

\maketitle

\begin{abstract}
We study the properties of a Rb magneto-optic trap loaded from a commercial 
getter source which provides a large flux of atoms for the trap along with 
the capability of rapid turn-off necessary for obtaining long trap 
lifetimes. We have studied the trap loading at two different values of 
background pressure to determine the cross-section for Rb--N$_2$ collisions 
to be $3.5(4) \times 10^{-14}$ cm$^2$ and that for Rb--Rb collisions to be 
of order $3 \times 10^{-13}$ cm$^2$. At a background pressure of $1.3 \times 
10^{-9}$ torr, we load more than $10^8$ atoms into the trap with a time 
constant of 3.3 s. The $1/e$ lifetime of trapped atoms is 13 s limited only 
by background collisions.
\end{abstract}

Laser cooling and trapping has now become a standard technique for atomic 
physics experiments whenever a cold dense sample of atoms is required. The 
workhorse in this field is the magneto-optic trap (MOT) \cite{RPC87}. For 
instance, a MOT is the starting point for laser cooling experiments 
involving Bose-Einstein condensation, cold collisions, optical lattices, 
precision spectroscopy, atomic fountain clocks, {\it etc} \cite{nobel97}. 
The source of atoms for the MOT is typically an oven inside the vacuum 
chamber containing a small amount of pure metal. The MOT is loaded in one of 
two ways: from a thermal beam slowed by a counter-propagating laser beam 
(Zeeman-slower or chirp slower), or from the slow atom tail of hot 
background vapor (so called vapor cell traps). In both cases, the atom 
density has to be quite high for the MOT to be loaded with a large number of 
atoms in a short time. But the high density usually limits the lifetime of 
the trap due to collisions with untrapped atoms unless the source can be 
turned off rapidly. This problem is solved in traps loaded with slowed 
atomic beams by using shutters to block the beam. In vapor cell traps, one 
solution is to use a double MOT, where the atoms are first loaded into a MOT 
located in a region of high vapor pressure, and then transferred through a 
differentially pumped line to a MOT in a region with much better vacuum 
where the lifetime is several orders of magnitude higher.

To address the same problem, we have been studying the loading of a MOT for 
Rb atoms from a commercial getter source \cite{SAES}. Rb getter sources 
release atomic Rb when resistively heated with a few A of current. They are 
inexpensive, easy to handle inside ultra-high vacuum (UHV) systems, and have 
the advantage that the supply of atoms can be rapidly switched off by 
turning down the heating current below a threshold value. Such sources have 
only recently started being used in laser cooling experiments 
\cite{WFG95,FGH98}. In Ref.\ 5, fast loading of a Rb MOT was achieved by 
placing the source within 30 mm of the trap center and operating it in a 
pulsed mode. By contrast, our source is placed about 90 mm away from the 
trap center and is operated continuously. We use the source as a thermal 
source much like the background vapor in vapor cell traps, but with the 
capability of rapid turn-off. Even though the stainless steel boat holding 
the source rises to a few hundred degrees C during operation, there are 
enough atoms within the velocity capture range of the laser cooling beams to 
load the MOT with more than $10^8$ atoms in a few seconds. To characterize 
the collisional loss mechanisms operating during the loading process, we 
have studied the properties of the MOT at different source currents and 
background pressures. From this, we extract the cross-section for Rb--N$_2$ 
and Rb--Rb collisions. In addition, we have measured the steady state trap 
population as a function of laser detuning and magnetic field gradient since 
optimal values of these parameters depend on the velocity distribution of Rb 
atoms emanating from the source. These results indicate that getter sources 
are compatible with fast loading of a large number of atoms into the MOT, 
along with a long trap lifetime after turn-off.

\section{Getter source details}

The getter source is a commercially available dispenser of alkali atoms 
designed for industrial applications. The source we use contains a Rb 
compound and a reducing agent enclosed in a stainless steel boat. The 
compound is stable at room temperature and the boat can be handled easily 
without fear of contamination. When a few A of current is passed through the 
boat, its temperature rises to several hundred degrees C. At these high 
temperatures, the Rb compound undergoes a reduction reaction and atomic Rb 
is released. Since the reaction is a threshold process, no Rb vapor is 
released below a threshold value of current, typically about 2.7 A.

We have installed the source in our vacuum system by attaching it to a UHV 
feedthrough using Cu-Be screw-type connectors. We have also tried spot 
welding the source to the feedthrough leads to reduce contact resistance, 
and find no significant difference in its performance. We prefer the   
screw-type connectors since this makes replacement of the source easier. 
Once inside the vacuum system, it is important to degas the source very well 
because this limits the ultimate base pressure in the chamber. We therefore 
operate the source at a current just below threshold for several hours 
during the initial bakeout of the system. When we first start using the 
source after bakeout, we find that no Rb is released even up to a current of 
5 A. To get it started, we need to raise the current to about 8 A for a 
short duration of 2--5 s. After several such high current pulses, the source 
becomes very reproducible and has a well defined threshold current of 2.7 A. 
We speculate that these pulses are essential to get rid of a surface crust 
that forms during storage and bakeout.

We have probed the Rb vapor density inside the vacuum system at different 
source currents using near-resonant laser light. Above the threshold current 
of 2.7 A, the fluorescence signal from the vapor increases rapidly with 
source current. At a current of 4 A, the vapor becomes optically dense. 
Around 4.5 A, the boat starts to glow red indicating a temperature near 
600$^{\rm o}$C. Above 8 A, so much Rb is released that the walls of the 
chamber get a metallic coating visible to the naked eye. We therefore load 
the MOT at source currents in the range of 3.0 to 3.8 A, which gives us an 
adequate flux of atoms without overloading the vacuum system. The fast   
turn-off capability of the source is seen in Fig.\ \ref{src_decay}, where we 
plot the decay of the background fluorescence after the current is turned 
off. The $1/e$ time constant is only about 3 s and is independent of source 
current in the wide range of 3 to 4 A. This indicates that the cooling of 
the source is very rapid and the time constant is probably limited by the 
pumping speed near the trapping region.

We find the source to be very robust. There is no noticeable change in the 
Rb flux at a given current even after a year of regular use. Though the 
supplier specifies a ``use before'' date of one year from the date of 
manufacture, we have installed it in our system even after three years 
without any problems. Contrary to what is reported in Ref.\ 5, the source 
does not appear to undergo ``irreversible contamination'' if the pressure 
rises by several orders of magnitude during initial degassing. Perhaps what 
they call irreversible contamination is similar to our observation that Rb 
atoms are not released even at currents of 5 A when the source is operated 
the first few times after bakeout. But, as mentioned earlier, this can be 
reversed by applying high current pulses to evaporate any surface layer that 
might have formed.

\section{Experimental details}

The experiments are done in a vacuum chamber consisting of a 100 mm O.D. 
stainless steel tube with ten intersecting 38 mm O.D. ports, as shown 
schematically in Fig.\ \ref{schematic}. Nine of the ten ports have 
commercial glass viewports for optical access, and one port holds the 
feedthrough for the getter source. The source is located about 90 mm from 
the center of the MOT. The chamber is pumped by a 300 $\ell$/s ion pump 
through a 100 mm O.D. port. The ion pump has a liquid nitrogen cooled    
cryo-panel to get better base pressure. Under normal conditions, the 
pressure inside the chamber is about $1\times 10^{-8}$ torr, which goes down 
to about $1\times 10^{-9}$ torr with the cryo-panel cooled to 77 K.

The MOT laser beams are derived from two home-built grating feedback 
stabilized diode lasers operating at 780 nm. The first laser is the cooling 
laser which is locked near the $5S_{1/2}, F=2 \rightarrow {5P_{3/2}}, F'=3$ 
cycling transition in $^{87}$Rb using saturated absorption spectroscopy in a 
room temperature vapor cell. The error signal is obtained by passing the 
laser beam through an acousto-optic modulator (AOM) and dithering the 
frequency shift in the AOM. The center frequency of the AOM is used to get 
variable detunings in the range of $-0.5 \Gamma$ to $-4\Gamma$, where 
$\Gamma = 2\pi \times 6.1$ MHz is the natural linewidth 
of the transition. Each of the 
three beams for the MOT is circularly polarized with a power of around 4 mW 
and a $1/e^2$ radius of 9 mm. The hyperfine repumping beam is derived from a 
second stabilized diode laser which is locked to the $5S_{1/2}, F=1 
\rightarrow {5P_{3/2}}, F'=2$ transition in a vapor cell. The error signal 
for this laser is obtained by modulating the diode injection current. The 
resulting variation of a few MHz in the repumping frequency is 
inconsequential for operation of the MOT. About 2 mW/cm$^2$ of repumping 
light is mixed with the cooling beams in a polarizing beamsplitter cube. The 
six laser beams intersect over a volume of 3 cm$^3$. A spherical quadrupole 
magnetic field is superposed on the intersection region using a pair of 
anti-Helmholtz coils (10 cm dia, 60 turns each) placed 10 cm apart. The 
field gradient at the center can be varied from 5 to 17 G/cm by varying the 
current through the coils in the range of 2 to 7 A.

The fluorescence from the trapped cloud of atoms is imaged on to both a CCD 
camera for viewing, and a calibrated silicon photodiode for quantitative 
measurements. To estimate the number of trapped atoms $N$, we fit the total 
optical power $P$ at the photodiode to the following expression:
\begin{equation}
\label{number}
P = \frac{hc}{\lambda} \frac{\Omega}{4\pi} N \gamma_{\rm sc}
\end{equation}
where $\Omega$ is the solid angle subtended by the detector at the cloud and 
$\gamma_{\rm sc}$ is the photon scattering rate given by:
\begin{equation}
\gamma_{\rm sc} = \frac{\Gamma}{2} \frac{I/I_0}{1 + I/I_0 + (2 \Delta / \Gamma)^2}
\end{equation}
with $I$ the total beam intensity from the six trapping beams, $I_0$ the 
saturation intensity for the transition (1.6 mW/cm$^2$), and $\Delta$ the 
detuning.

\section{Results}
We have studied the loading characteristics of the MOT by measuring the trap 
fluorescence as a function of time after the laser beams are turned on. The 
source is turned on a few minutes before the light beams so that the flux of 
Rb atoms reaches a steady state. The loading of atoms into the trap is 
determined by a balance between the capture rate and the loss rate. This 
results in the following rate equation:
\begin{equation} 
\label{rateeq} 
\frac{dN}{dt} = R - \frac{N}{\tau} 
\end{equation}
where $N$ is the number of atoms in the trap, $R$ is the rate at which atoms 
are captured from the background vapor, and $\tau$ is the loading time 
constant determined by losses due to collisions \cite{REI65}. The loss rate 
depends both on collisions with background atoms as well collisions with 
hot, untrapped Rb atoms:
\begin{equation} 
\label{tau} 
\frac{1}{\tau} = n_{\rm b} \sigma_{\rm b} \bar{v}_{\rm b} 
+ n_{\rm Rb} \sigma_{\rm Rb} \bar{v}_{\rm Rb}  
\end{equation}
where $n$ is the density of scattering particles, $\sigma$ is the 
collisional scattering cross section and $\bar{v}$ is the average velocity. 
The subscript indicates whether the scattering is from the background (b) or 
from the Rb source. The solution to Eq.\ \ref{rateeq} is an exponential 
growth in the number of trapped atoms:
\begin{equation}
\label{ns} 
N = N_s [1- \exp(-t/ \tau)]
\end{equation}
with a steady state value of $N_s=R\tau$. Since the trap fluorescence is 
directly related to the number of trapped atoms (from Eq.\ \ref{number}), 
the photodetector signal after the laser beams are turned on is of the above 
form. The trap lifetime $\tau_{\rm d}$ after the source is shut off is 
determined solely by losses due to collisions with non-Rb 
background atoms. Therefore, the 
photodetector signal after the source is turned off shows an exponential 
decay with a time constant independent of the steady state population. A 
typical trap loading and decay curve taken at a source current of 3.8 A is 
shown in Fig.\ \ref{loading}.

To estimate the values of $\sigma_{\rm b}$ and $\sigma_{\rm Rb}$ in Eq.\ 
\ref{tau}, we have studied the trap loading and decay properties at 
different source currents and two values of background pressure, namely 
$1.5\times 10^{-8}$ torr and $1.3\times 10^{-9}$ torr. The results are 
listed in Table \ref{data}. Several features of the data are to be noted. 
First of all, at the higher pressure the loading time $\tau$ is 1 s 
independent of the source current, indicating that background gas collisions 
dominate the loss rate. Next, at the lower pressure, $\tau$ goes down from 
4.6 s at 3.0 A to 3.3 s at 3.8 A as the flux of atoms from the source 
increases. The steady state trap population $N_s$ increases correspondingly 
from $1.1\times 10^7$ to $1.1\times 10^8$. Finally, the trap lifetime 
$\tau_{\rm d}$ is independent of source current (and $N_s$) at both values 
of pressure. This is consistent with the fact that the lifetime depends only 
on background gas collisions. However, at the higher pressure, one would 
expect that background collisions would limit the lifetime to a value around 
1 s in order to be consistent with the background limited loading time of 1 
s, but the measured value is actually 5.5 s. This difference arises because 
the source has a decay constant of 3 s after turn-off (see Fig.\ 
\ref{src_decay}) and the trap continues to be loaded even after the source 
is off. At lower pressure, the measured lifetime of the trap is 13.3 s; here 
the 3 s decay of the source can be neglected and the turn off can be assumed 
to be instantaneous.

We use the values of $\tau$ for different pressures at the {\it same source 
current} to eliminate the effect of the Rb term in Eq.\ \ref{tau} and obtain 
the cross section for background gas collisions from the following 
expression:
\begin{equation} 
\label{sigmab} 
\frac{1}{\left( \tau \right)_{\rm h}} - 
\frac{1}{\left( \tau \right)_{\rm l}}
= \left[ ( n_{\rm b} )_{\rm h} - ( n_{\rm b} )_{\rm l} \right] 
\sigma_{\rm b} \bar{v}_{\rm b}
\end{equation}
where the subscripts h and l indicate higher and lower pressure 
respectively. Assuming the background is predominantly nitrogen, which seems 
reasonable given the considerable drop in pressure when the cryo-panel is 
cooled to 77 K, we can calculate the values of $n_{\rm b}$ and $\bar{v}_{\rm 
b}$ at room temperature for a given pressure. The above equation then yields 
the value of $\sigma_{\rm b}$. The average value of $\sigma_{\rm b}$ 
obtained from the values of $\tau$ at source currents of 3.0 A and 3.3 A is 
$3.5(4) \times 10^{-14}$ cm$^2$. We can also obtain the value of 
$\sigma_{\rm b}$ from the 13.3 s trap lifetime at lower pressure assuming 
that the loss is solely due to background gas collisions. This yields a 
value of $3.8 \times 10^{-14}$ cm$^2$, consistent with the above result. 
These values are also comparable to the cross section of $3.3 \times    
10^{-14}$ cm$^2$ for Na--N$_2$ collisions reported in Ref.\ 7.

Using the above value of $\sigma_{\rm b}$, we obtain the value of 
$\sigma_{\rm Rb}$ in the following manner. We first calculate the capture 
rate $R$ by assuming, for simplicity, that all atoms entering the trap 
volume with a velocity less than a critical velocity $v_c$ are slowed by the 
laser beams and captured. If we further assume that the background vapor has 
a thermal distribution, the number of such atoms per unit time can be 
calculated from kinetic theory \cite{REI65}, which yields:
\begin{equation} 
\label{rate} 
R = \frac{2}{\pi^2} n_{\rm Rb} A \frac{v_c^4}{\bar{v}_{\rm Rb}^3} 
\end{equation}
where $A$ is the surface area of the trapping region determined by the 
overlap of the laser beams (10 cm$^2$ in our trap). From a one-dimensional 
laser cooling model, $v_c$ is the velocity at which the Doppler shift takes 
the atom out of resonance by one linewidth, therefore it is given by $(| 
\Delta | + \Gamma) \lambda /2\pi$, about 14 m/s for a typical detuning of  
$-2\Gamma$. Using Eq.\ \ref{tau} for $\tau$, and Eq.\ \ref{rate} for $R$, we 
obtain:
\begin{equation}
\label{rsigma} 
\sigma_{\rm Rb} = \frac{2}{\pi^2} \frac{A}{N_s}
\left( \frac{v_c}{\bar{v}_{\rm Rb}} \right)^4
\left[ 1- \tau \left( n_{\rm b} \sigma_{\rm b} \bar{v}_{\rm b} \right) 
\right]
\end{equation}
which can be used to calculate $\sigma_{\rm Rb}$ if $\bar{v}_{\rm Rb}$ is 
known. Assuming a value of 355 m/s for $\bar{v}_{\rm Rb}$ (corresponding to 
a temperature of 250$^{\rm o}$C), we determine a value of $3 \times 10^{-
13}$ cm$^2$ for $\sigma_{\rm Rb}$ from the data at 3.0 A. This value 
compares well with the following theoretical estimate for the scattering 
cross section which assumes that the dominant scattering mechanism is 
resonant dipole-dipole scattering \cite{SCF92}:
\begin{equation} 
\label{sigma} 
\sigma_{\rm Rb} = \pi \left( \frac{4 C_3}{m_{\rm Rb} v_{\rm esc} 
\bar{v}_{\rm Rb} } \right)^{2/3} 
\end{equation}
where $C_3 = 5.8\times 10^{-48}$ Jm$^3$ for Rb. Using $v_c = 14$ m/s for 
$v_{\rm esc}$ and 355 m/s for $\bar{v}_{\rm Rb}$ yields a value of $3.2 
\times 10^{-13}$ cm$^2$ for $\sigma_{\rm Rb}$. Therefore, our assumption 
that the temperature of the vapor is around 250$^{\rm o}$C seems reasonable. 
At higher currents of 3.3 A and 3.8 A, the measured values of $\tau$ lead to 
smaller estimates of $\sigma_{\rm Rb}$. This is probably because nonlinear 
loss terms \cite{APE94} become important at high atom densities and Eq.\ 
\ref{rateeq} is no longer valid. It is difficult to extract better estimates 
of the Rb--Rb cross section without a more complete knowledge of the 
distribution of atoms coming out of the source at different currents.

The velocity distribution of atoms emanating from the source is also 
important in determining optimal values of detuning and magnetic field 
gradient for the MOT. This is because the detuning determines the capture 
velocity, and, as we will see below, the field gradient determines the 
effective slowing distance. We have therefore studied the steady state trap 
population $N_s$ as a function of these parameters. The dependence on 
detuning is shown in Fig.\ \ref{Ns_det}. The number of trapped atoms is 
maximized at a detuning of $-2.5 \Gamma$ for a field gradient 
of 10 G/cm. The dependence can be understood from the following physical 
picture. At low values of detuning, the critical velocity $v_c$ for slowing 
is small, and fewer atoms are captured since $R$ varies as $v_c^{4}$ (see 
Eq.\ \ref{rate}): for example, when $\Delta = -\Gamma$, $v_c$ is only 9.5 
m/s. At high values of detuning, even though $v_c$ increases, the slowing 
distance also increases and atoms are not slowed within the 9 mm size of the 
laser beams. Thus, when $\Delta = -4 \Gamma$, $v_c$ increases to 24 m/s but 
the slowing distance is about 11 mm. The optimal value of detuning that we 
measure is similar to what is observed in a MOT loaded from room temperature 
vapor \cite{LSW92}.

The dependence of $N_s$ on the magnetic field gradient for a fixed detuning 
is shown in Fig.\ \ref{Ns_mag}. The trap population is again maximized at an 
optimum value of field gradient. The trend can be understood as follows. At 
low values of the field gradient, the trap is shallow and atoms leave the 
trap easily. At high values of the gradient, the Zeeman shift, which 
increases rapidly away from the trap center, makes laser cooling ineffective 
because it changes the effective detuning to a smaller value that is less 
optimal for cooling. For example, the Zeeman shift for $F=2 \rightarrow 
F'=3$ transitions driven by circularly polarized light is 1.4 MHz/G. If the 
detuning is $-2 \Gamma$ and the field gradient is 17 G/cm, the atoms become 
resonant with the laser at a distance of 5 mm from the trap center and atoms 
are not cooled beyond this point. The optimal value of field gradient we 
measure is comparable to values obtained in vapor cell MOTs \cite{LSW92}. 
The consistency of optimal values of detuning and field gradient with 
results from vapor cell traps indicate that the velocity distribution 
emanating from the source is thermal.

\section{Conclusions}
We have demonstrated that a MOT can be loaded efficiently from a getter 
source. The source combines the advantages of providing a large flux of 
atoms that can be captured by the MOT with a fast turn-off necessary for 
getting a long trap lifetime. The loading rate is determined both by 
collisions with background atoms and hot untrapped Rb atoms. We have 
estimated the cross section for Rb--N$_2$ and Rb--Rb collisions by measuring 
the trap properties at two values of background pressure. The value of 
$\sigma_{\rm b}$ is $3.5(4) \times 10^{-14}$ cm$^2$, comparable to a value 
reported for Na--N$_2$ collisions, and the value of $\sigma_{\rm Rb}$ is of 
order $3 \times 10^{-13}$ cm$^2$ consistent with a theoretical estimate 
based on resonant dipole-dipole scattering. At a pressure of $1.3 \times 
10^{-9}$ torr, we load more than $10^8$ atoms into the MOT with a time 
constant of 3.3 s. At a background pressure of $1.5\times 10^{-8}$ torr, the 
loading time constant is 1 s independent of the source current, indicating 
that collisions with the background are dominant. We have also measured 
optimal values of detuning and magnetic field gradient which maximize the 
trap population. These values are consistent with observations in vapor cell 
MOTs, indicating that the velocity distribution of Rb atoms from the source 
is thermal. The trap lifetime at a pressure of $1.3 \times 10^{-9}$ torr is 
13 s and is limited only by the background pressure since the supply of Rb 
atoms stops almost instantaneously. We are in the process of changing over 
to a new vacuum system with a Ti-sublimation pump that should give us base 
pressure in the range of $10^{-11}$ torr where we estimate that the trap 
lifetime will be a few minutes.

\acknowledgements
We gratefully acknowledge the help of machinists at the Tool Room of the 
Centre for Product Design and Manufacturing, Indian Institute of Science, 
for fabricating the optical mounts and diode laser system hardware. This 
work was supported by a research grant from the Department of Science and 
Technology, Government of India. One of us (AW) acknowledges financial 
support from a CSIR post-doctoral fellowship.

\begin{figure}
\caption{
Decay of atoms from the source. The photodetector signal measures the 
fluorescence from Rb atoms in the background vapor. The source current is 
turned off at $t=0$ after which the fluorescence decays exponentially with a 
time constant of about 3 seconds.
}
\label{src_decay}
\end{figure}

\begin{figure}
\caption{
Schematic of the experiment. The chamber is made of stainless steel and has 
nine optical ports. Two of the three pairs of counter-propagating laser 
beams are shown, a third pair is along the magnetic field axis. The Rb 
getter source is located about 9 cm from the center of the trap.
}
\label{schematic}
\end{figure}

\begin{figure}
\caption{
Loading and decay of the MOT. The plot shows a typical measurement of the 
trap fluorescence as a function of time. The photodetector signal is 
proportional to the number of trapped atoms. The trapping beams are turned 
on at $t=0$, after which the signal shows an exponential growth with a time 
constant of about 4 s (see Eq.\ \ref{ns} in the text). The signal decays 
exponentially after the source is turned off with a $1/e$ lifetime of about 
13 s. The data were taken with a source current of 3.8 A.
}
\label{loading}
\end{figure}

\begin{figure}
\caption{
$N_s$ {\it vs} detuning. The plot shows the steady state number of trapped 
atoms as a function of detuning for a constant magnetic field gradient of 10 
G/cm. The number is sharply peaked at a detuning near $-2.5 \Gamma$. The 
error bars represent statistical errors and do not include systematic errors 
in scaling the photodetector signal to number of atoms. Such systematic 
errors, however,
will not affect the observed trend.
}
\label{Ns_det}
\end{figure}

\begin{figure}
\caption{
$N_s$ {\it vs} magnetic field gradient. The plot shows the steady state 
number of trapped atoms as a function of axial field gradient for a fixed 
detuning of $-2 \Gamma$. The number shows a broad maximum at values of 
$dB/dz$ around 12 G/cm. The error bars again represent only statistical 
errors, but as for Fig.\ \ref{Ns_det}, systematic errors are not important 
for the trend.
}
\label{Ns_mag}
\end{figure}

\mediumtext
\begin{table}
\caption{ 
The table lists the values of $\tau$, $N_s$, and $\tau_{\rm d}$ measured at 
different source currents at two values of background pressure. All the data 
were taken with an intensity of about 4 mW/cm$^2$ in each laser beam, a 
detuning of $-2 \Gamma$ and a magnetic field gradient of 10 G/cm.
\label{table1}}
\begin{tabular}{ccccccc}
 & \multicolumn{3}{c}{Pressure $=1.5 \times 10^{-8}$ torr} &
\multicolumn{3}{c}{Pressure $=1.3 \times 10^{-9}$ torr} \\
\cline{2-4} \cline{5-7}
Source current (A) & $\tau$ (s) & $N_s$ & $\tau_{\rm d}$ (s) & $\tau$ (s) & 
$N_s$ & $\tau_{\rm d}$ (s) \\
\tableline
3.0 & 1.0(2) \hspace*{0.7mm} & $3.4\times 10^6$ & 5.5(4) & 4.6(2) & 
$1.1\times 10^7$ & \hspace*{0.7mm} 13.3(11) \\
3.3 & 1.06(2)  & $1.1\times 10^7$ & 5.5(3) & 4.2(2) & $3.5\times 10^7$ & 
13.2(7) \\
3.8 & 1.05(1)  & $5.0\times 10^7$ & 5.3(3) & 3.3(1) & $1.1\times 10^8$ & 
13.3(5) \\
\end{tabular}
\label{data}
\end{table}


\begin{references}

\bibitem{RPC87}
E.~L. Raab {\it et~al.}, Phys. Rev. Lett. {\bf 59},  2631  (1987).

\bibitem{nobel97}
A good review of laser cooling and trapping experiments is contained in the 
Nobel Prize lectures:
S. Chu, Rev. Mod. Phys. {\bf 70},  685  (1998),
C.~N. Cohen-Tannoudji, {\it ibid.\ }{\bf 70}, 707 (1998); and 
W.~D. Phillips, {\it ibid.\ }{\bf 70}, 721 (1998).

\bibitem{SAES}
SAES Getters S.p.A., Viale Italia 77-1-20020, Lainate (MI), Italy.

\bibitem{WFG95}
C. Wieman, G. Flowers, and S. Gilbert, Am. J. Phys. {\bf 63},  317  (1995).

\bibitem{FGH98}
J. Fortagh, A. Grossmann, T.~W. H\"ansch, and C. Zimmermann, J. Appl. Phys.
  {\bf 84},  6499  (1998).

\bibitem{REI65}
F. Reif, {\em Fundamentals of Statistical and Thermal Physics} (McGraw-Hill
  Inc., New York, 1965).

\bibitem{PCB88}
M. Prentiss {\it et~al.}, Opt. Lett. {\bf 13},  452  (1988).

\bibitem{SCF92}
A.~M. Steane, M. Chowdhury, and C.~J. Foot, J. Opt. Soc. Am. B {\bf 9},  
2142
  (1992).

\bibitem{APE94}
M.~H. Anderson, W. Petrich, J.~R. Ensher, and E.~A. Cornell, Phys. Rev. A 
{\bf
  50},  R3597  (1994).

\bibitem{LSW92}
K. Lindquist, M. Stephens, and C. Wieman, Phys. Rev. A {\bf 46},  4082  
(1992).

\end{references}
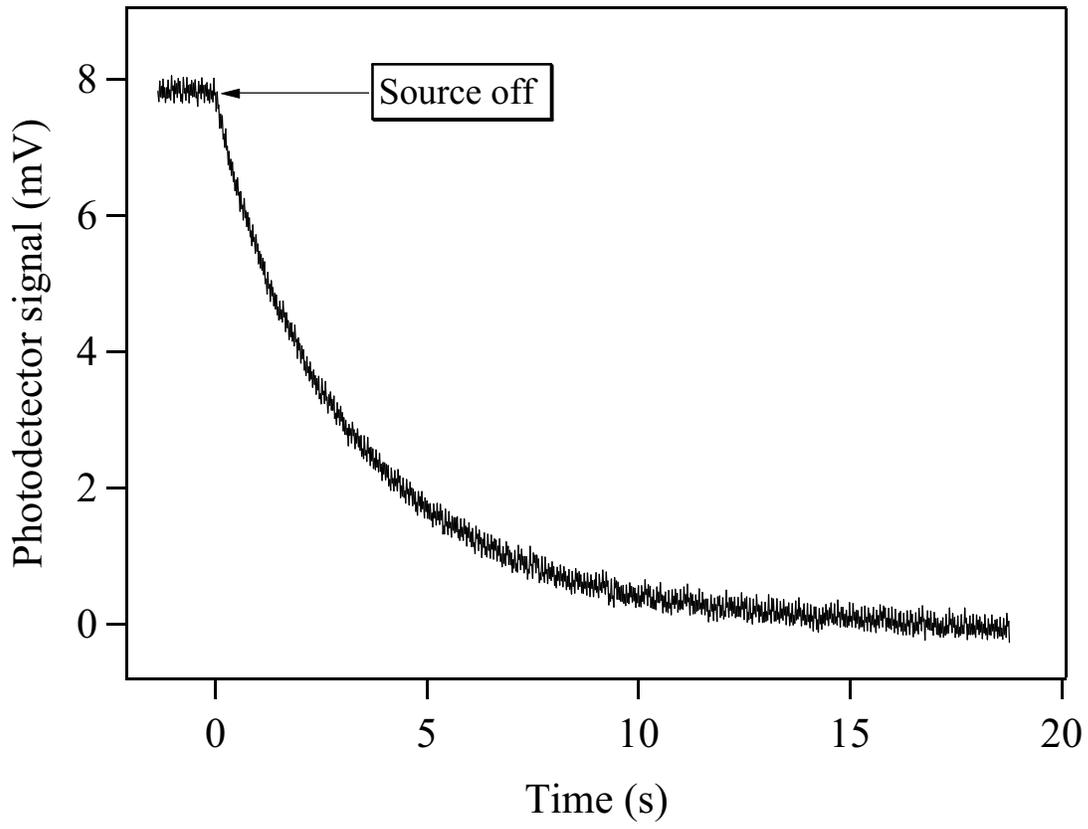
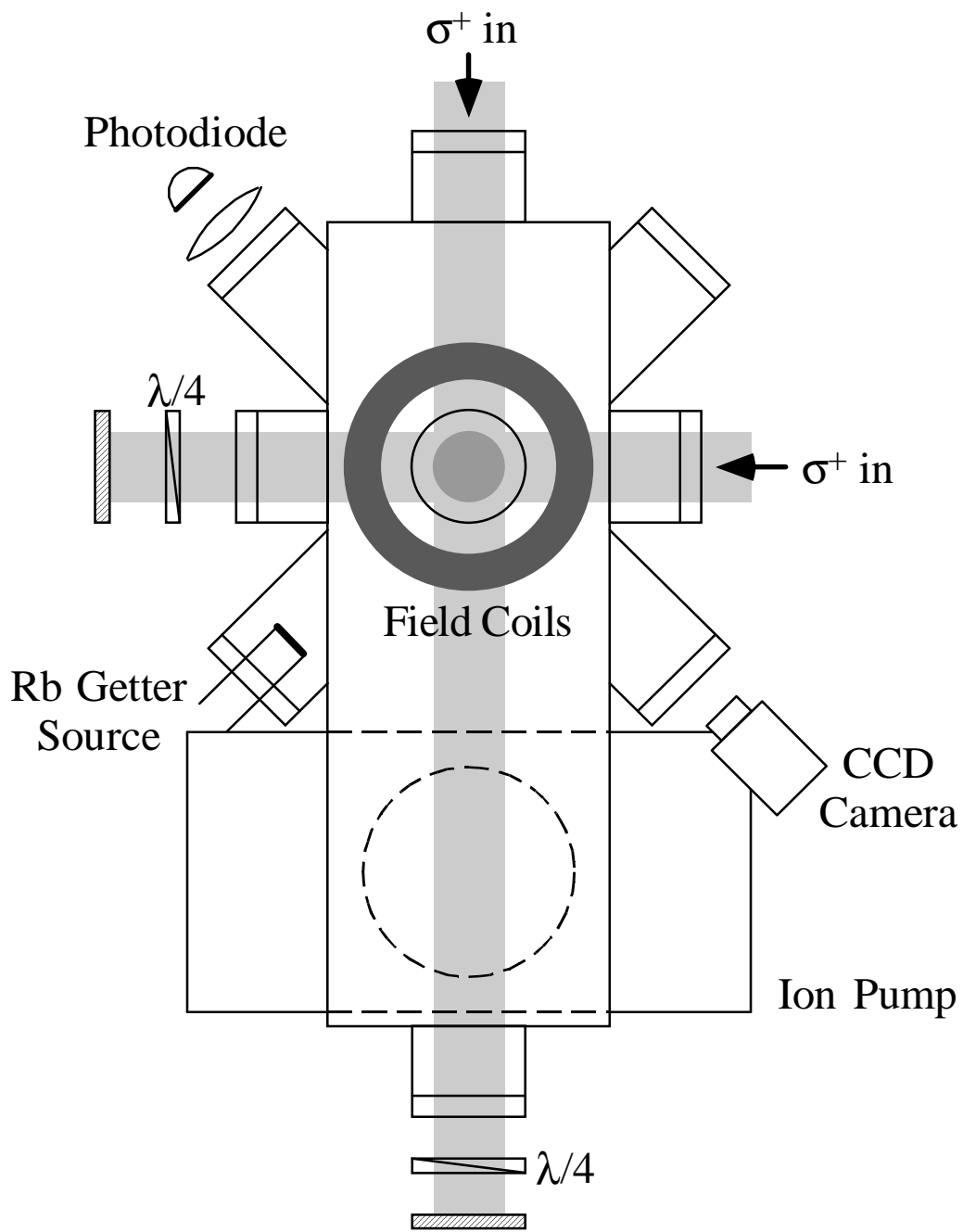
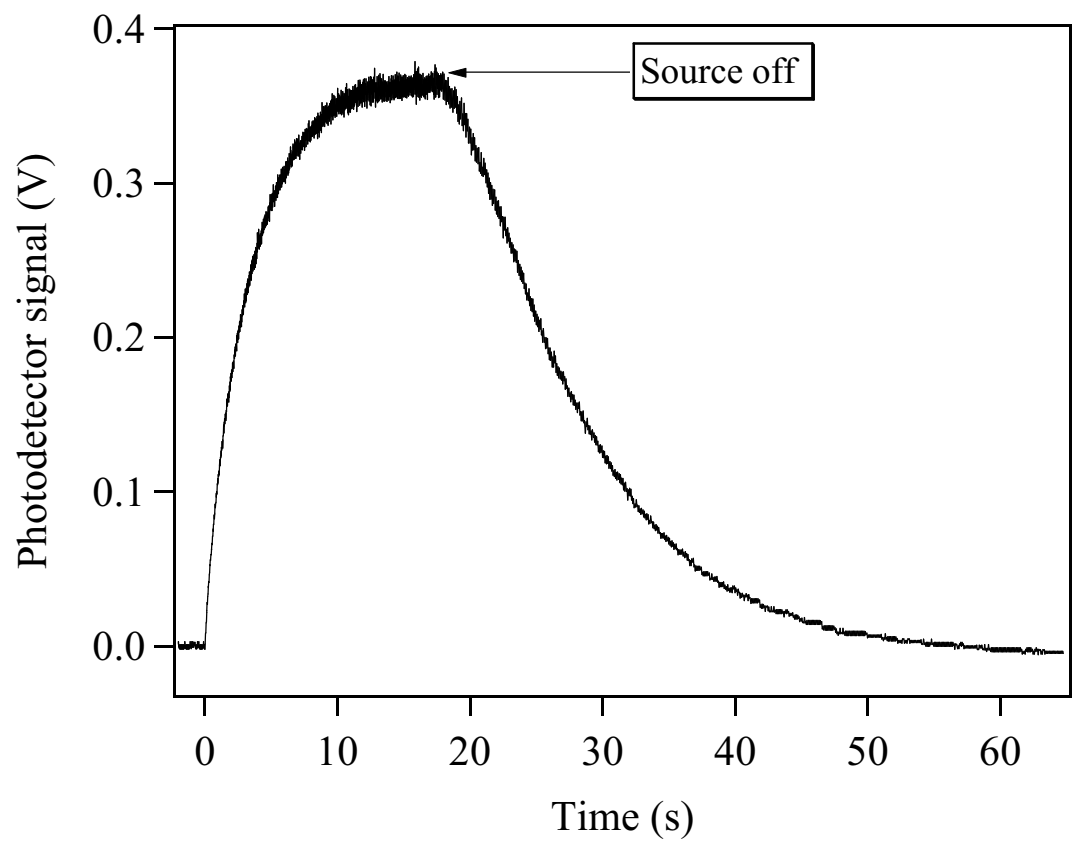
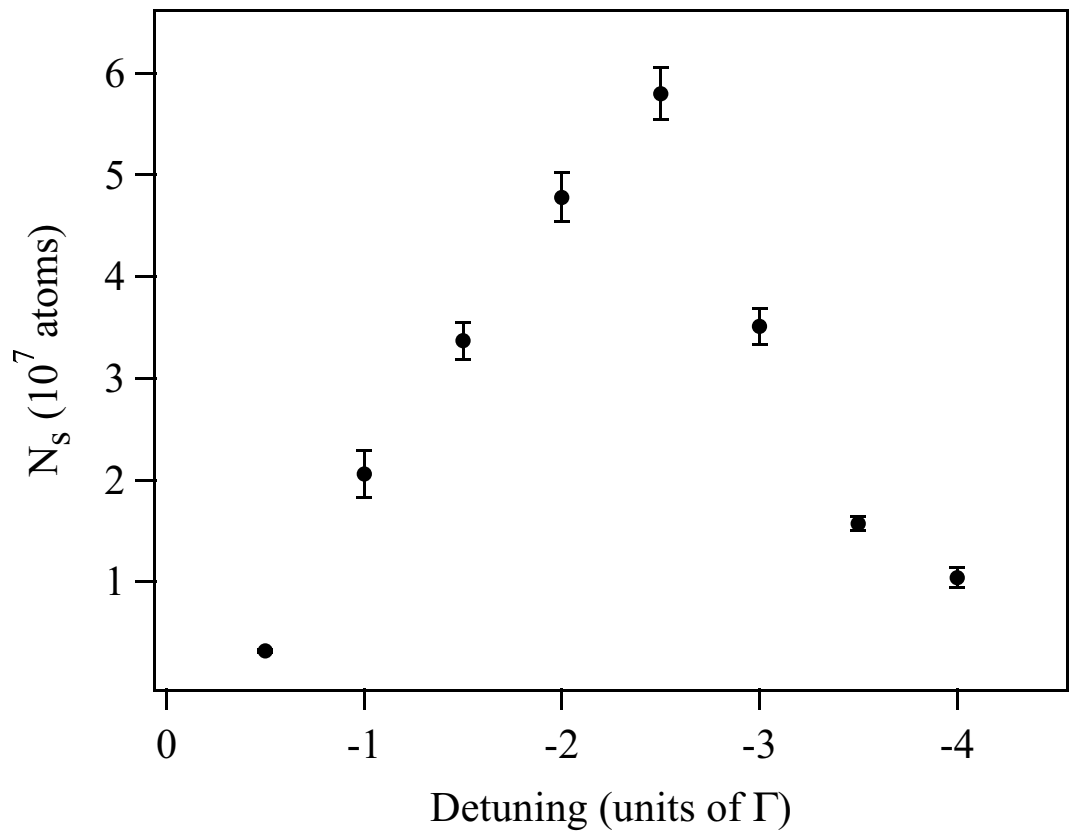
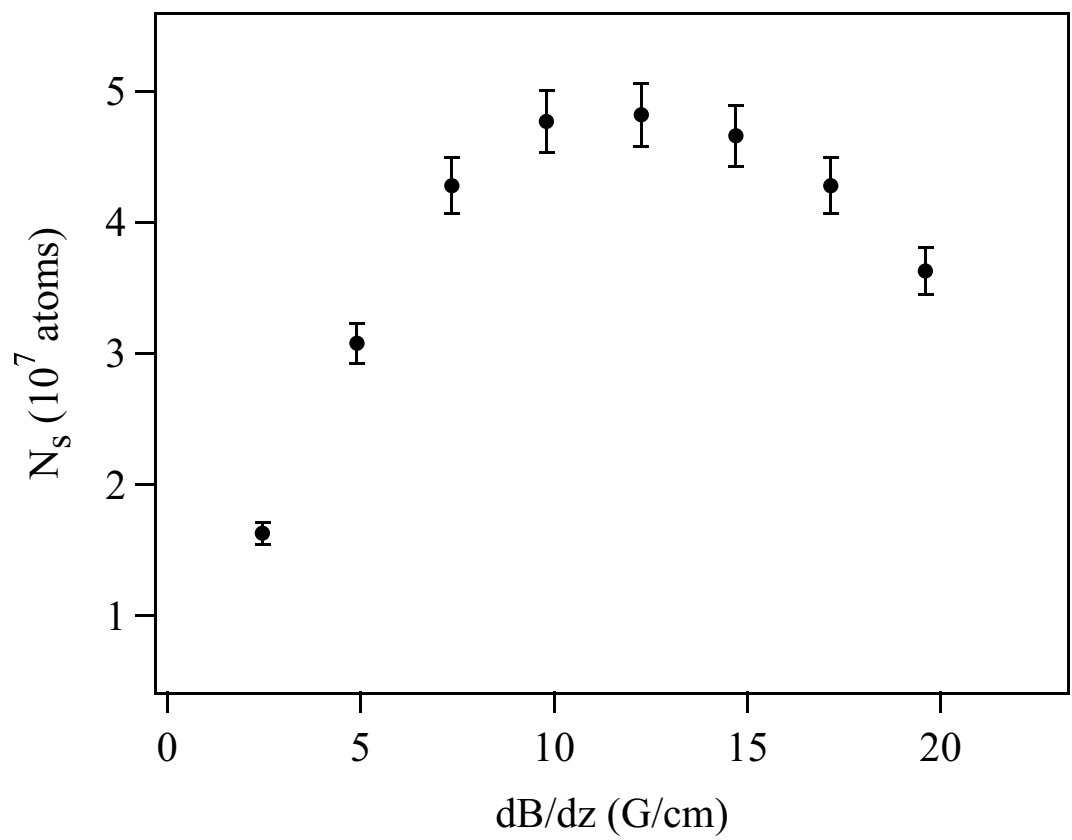
\end{document}